\documentclass[a4paper]{article}

\usepackage{INTERSPEECH2019}
\usepackage{amsmath,graphicx,amssymb,xcolor,multirow}
\usepackage[ruled]{algorithm2e}
\usepackage{epstopdf}
\usepackage{xcolor}

\def\v{{\mathbf v}}

\def\v#1{\mathbf{#1}}
\def\l#1{\mathcal{#1}}

\title{Compression of Acoustic Event Detection Models With Quantized Distillation}
\name{Bowen Shi\textsuperscript{1*}\thanks{*work done at Amazon}, Ming Sun\textsuperscript{2}, Chieh-Chi Kao\textsuperscript{2}, Viktor Rozgic\textsuperscript{2}, Spyros Matsoukas\textsuperscript{2}, Chao Wang\textsuperscript{2}}
\address{\textsuperscript{1}Toyota Technological Institute at Chicago\\
  \textsuperscript{2}Amazon Alexa Research}
\email{bshi@ttic.edu, \{mingsun,chiehchi,rozgicv,matsouka,wngcha\}@amazon.com}

\begin{document}

\maketitle
\begin{abstract}
  Acoustic Event Detection (AED), aiming at detecting categories of events based on audio signals, has found application in many intelligent systems. Recently deep neural network significantly advances this field and reduces detection errors to a large scale. However how to efficiently execute deep models in AED has received much less attention. Meanwhile state-of-the-art AED models are based on large deep models, which are computational demanding and challenging to deploy on devices with constrained computational resources.
  In this paper, we present a simple yet effective compression approach which jointly leverages \emph{knowledge distillation} and \emph{quantization} to compress larger network (teacher model) into compact network (student model). Experimental results show proposed technique not only lowers error rate of original compact network by $15\%$ through distillation but also further reduces its model size to a large extent ($2\%$ of teacher, $12\%$ of full-precision student) through quantization. 
\end{abstract}


\section{Introduction}

Acoustic event detection (AED), the task of detecting the occurrence of certain events based on audio streams, can be widely applied in many scenarios. In surveillance systems, audio is used either independently or in conjunction with visual modality for scene analysis. For example, \cite{road_surveillance} applies AED model to detect hazardous events and perilous road situations. AED also plays an essential role in enabling widespread voice-based virtual assistants (Amazon Alexa, Apple Siri, and Google Assistant etc) to better understand surrounding environments, beyond only listening to human speech. It can also enhance human-computer interaction by providing context-related information \cite{Shah2018ACL}. 

Most works on AED only focus on lowering detection error rate while studies on reducing inference latency are much fewer.  Models with high accuracies are often of deep architectures composed of multiple convolutional and recurrent layers \cite{many_cnns, rcnn, cnn_augment, cakir2017}, thus being computation exhaustive.
This raises a concern for the use of AED models in resource-constraint scenarios where computation resource and memory space are limited (e.g. mobile devices). 

In this paper, we look into model compression for AED. The objective is to reduce model size and speed up computation while preserve accuracy. Our compression approach, called \emph{quantized distillation}, is based on knowledge distillation and quantization. We mainly focus on using shallow recurrent neural network (RNN) as our student model and very deep Convolutional Neural Network (CNN) as the teacher model respectively due to their computational efficiency and high accuracy. Reason of this choice will be detailed in section \ref{sec:method}. 
   Nevertheless our compression technique is not restricted to specific model architectures and can be directly applied on other AED models.

\section{Related work}
\label{sec:relate_work}
Neural network compression has been well explored in broad context. Knowledge distillation \cite{kd} is a commonly used technique for model compression, which consists of training a compact student network with distilled knowledge from a large teacher network. Knowledge distillation has been widely applied in various domains, including automatic speech recognition (ASR) \cite{kd_asr_1,kd_asr_2,kd_asr_3}, visual object detection \cite{kd_obj}. Network quantization, another category of compression technique, refers to compressing the original network by reducing number of bits required to represent its weights. Quantization methods have been studied from the perspective of different model architectures \cite{quant_cnn,quant_rnn} as well the specific applications (e.g. ASR \cite{quant_asr}, machine translation \cite{kd_qt}). Besides, low rank factorization \cite{low_rank_1, low_rank_2, kw_1, kw_2} is also commonly used in model compression, which estimates informative parameters by using matrix decomposition. Individual compression techniques can also be combined to achieve larger compression rate. For instance, \cite{kd_asr_3} evaluates the combination of low-rank factorization, knowledge distillation and network pruning in ASR. Specifically in AED, \cite{alex_bn} investigates compression of CNNs and their method is on simplification of architectures by introducing bottleneck layers and global pooling. \cite{aed_svd_quant} combines quantization and low-rank matrix factorization technique to compress multi-layer recurrent neural network. In \cite{aed_kd} knowledge distillation is applied to train CNNs of small footprint.

This paper focuses on jointly applying knowledge distillation and quantization to AED models. Similarly, \cite{kd_qt} studies the grouping effect of above two techniques in image classification and machine translation . The distinctions between ours and \cite{kd_qt} are mainly in following two aspects. (1). We study this approach in AED. Specifically in this domain, we observe deep convolutional models outperform simple recurrent networks by large margin. We will show later in section \ref{sec:kd_effect} that the performance gap is largely related to the model architecture instead of just model size (e.g., depth, number of units per layer). The inherent problem specific to certain architecture, such as gradient vanishing in RNN when sequence is long, may make knowledge distillation less effective. Whether student model can still benefit from teacher model even when two architectures greatly differ is the question we address in this paper. In tasks studied in \cite{kd_qt}, teacher and student models only differ in size within same task.
 (2). We study the quantization scheme where both inputs and parameters are quantized into integers while \cite{kd_qt} focuses mainly on quantizing parameters. Our quantization scheme not only allows savings of storage but also accelerates computation. Under this scheme, quantization in training is more important to prevent loss of accuracy, which will be shown in section \ref{sec:analysis_qt}.

\section{Methods}
\label{sec:method}
We start by formulating the multi-class acoustic event detection problem. Given an audio signal $I$ (e.g. log mel-filter bank energies (LFBEs)), the task is to train a model $\v f$ to predict a multi-hot vector $\v y\in\{0, 1\}^C$, with $C$ being the size of event set $\l E$, and $y_c$ being a binary indicator whether event $c$ is present in $I$. Note the prediction $\v f(I)$ is not a distribution over event set $\l E$ since multiple events can occur simultaneously in $I$. We denote $\l D_L=\{(I, \v y)\}$ as the labeled dataset. Model $\v f$ is trained using cross-entropy loss (see equation \ref{eq:sup_loss}), where $w_c$ is the penalty of mis-classification for positive data of class $c$. $w_c$ should be tuned to balance losses computed from positive 
and negative instances.

\begin{equation}
\centering
\label{eq:sup_loss}
L_{ce}=-\displaystyle\sum_{(I, \v y)\in \l D_L}\displaystyle\sum_{c=1}^C\{w_c y_c\log f_c(I)+(1-y_c)\log(1-f_c(I))\}
\end{equation}

\textbf{Knowledge Distillation}
Classic knowledge distillation \cite{kd} requires training a teacher model first, which is often a large network. A student model is trained with knowledge distillation loss which utilizes the logits outputs from the teacher model as soft-targets. In our multi-class setting, loss  defined in equation \ref{eq:kd_loss} is used to train the student model $\mathbf{m}^s$.

\begin{equation}
\label{eq:kd_loss}
\centering
\begin{split}
& L_{kd}=\displaystyle\sum_{(I, \v y)\in\l D_{L}}\{\alpha T^2 l(I, \v y^t(I; T))+(1-\alpha)l(I, \v y)\} \\
  & l(I, \v y^\prime) = \displaystyle\sum_{c=1}^C\{w_c y^{\prime}_c\log \mathbf{m}^s_c(I)+(1-y^{\prime}_c)\log(1-\mathbf{m}^s_c(I))\} \\
& \v y^t(I;T) = \frac{1}{1+\exp(-\frac{\v m^t(\mathbf{I})}{T})} \\
\end{split}
\end{equation}

where $\v m^t$ and $\v m^s$ are teacher and student model which takes in acoustic signal $\v I$ and output logits in different event categories, $\v y^\prime$ denotes probability of either ground-truth ($\v y$) or that given by teacher model ($\v y^t(I;T)$). $T$ and $\alpha$ are hyperparameters controlling respectively the softness of teacher logits $\v z$ and relative weight of distillation loss. 

\textbf{Quantized Distillation}
Based on classic knowledge distillation, we further quantize the student model both during training and inference time to introduce more compression effect, which we refer to as \emph{quantized distillation} throughout the paper. Quantization consists of representing floating-point values with n-bit integers ($n<32$). A common quantization process include following steps: (1). scaling: which normalizes vectors of arbitrary range to values in $[0, 1]$, (2). quantizing: which rounds scaled values to quantized values $[0, 1]$, (3). recovering: which scales quantized values in $[0, 1]$ back to original range. The above process is formulated as equation \ref{eq:quantization}, where $\v V$ and $\mathbf{V}^q$ are parameter before and after quantizing.

\begin{equation}
\label{eq:quantization}
\begin{split}
& \hat{\v V} = \frac{\v V-\beta}{\alpha} \\
& \hat{\v V}^q=\hat{Q}_{n}(\hat{\v V}) = \frac{[\hat{\v V}(2^n-1)]}{2^n-1} \\
& \mathbf{V}^q = \alpha\hat{\v V}^q+\beta \\ 
& \alpha = \max_{i}{V_i}-\min_i{V_i},\ \beta=\min_i{V_i} \\
\end{split}
\end{equation}

As the quantization function ($\hat{Q}_n$ in equation \ref{eq:quantization}) is discrete, its gradient is almost zero everywhere. To solve this problem, we apply straight-through estimator \cite{straight_through} to approximate the gradient computation regarding $\hat{\v V}$. The forward and backward pass are given as equation \ref{eq:forward_backward}.

\begin{equation}
\label{eq:forward_backward}
\begin{split}
\bf forward: & \hat{\v V}^q = \hat{Q}_{n}(\hat{\v V}) \\
\bf backward: & \frac{\partial l}{\partial\hat{\v V}} = \frac{\partial l}{\partial\hat{\v V}^q} \\
\end{split}
\end{equation}

During training time, we train a quantized student model by distilling from a full-precision teacher model through minimizing loss of equation \ref{eq:kd_loss}. At test time, only the quantized student model is used for inference.

\textbf{Choice of Model}
In the scope of this paper, the student model is based on RNN and teacher model is on CNN. Across several benchmarks \cite{many_cnns,cnn_augment,rcnn} in AED, CNNs including classic convolutional model and convolutional architecture augmented with additional recurrent layers has achieved much better detection accuracy. Those CNNs are usually very deep in consideration of accuracy. For example, the best performing model in \cite{many_cnns} is ResNet \cite{resnet} with 50 layers. Compared to CNNs, RNN has folllowing advantages:
 (1). It is more compact and induces much less computation compared to a deep CNN (see table \ref{tab:model_size} in experimental section for detailed comparison) (2). For CNN, entire sequence of raw input and its sub-sampled feature sequence of each intermediate layer have to be stored in memory during forward computation. The memory overhead is huge particularly when CNN is deep (e.g., ResNet-50). This arouses a concern for its deployment on devices with limited computational resources. (3). RNN is naturally suited for on-line setting where sequence length is undetermined beforehand, which is often the case in real-life applications of AED.

\textbf{LSTM Quantization} We choose LSTM as our student RNN model and the remaining part of this section will be devoted to how to quantize LSTM in detail. The structure of LSTM is shown as equation \ref{eq:lstm}, where $x_t$, $h_t$, $C_t$ are input, hidden state, cell state of LSTM at timestep $t$. $\{W_f, W_i, W_c, W_o\}$, $\{b_f, b_i, b_c, b_o\}$ are weight and bias parameters to be trained. 

\begin{equation}
  \label{eq:lstm}
  \begin{split}
    & f_t = \sigma(W_f\cdot [h_{t-1}, x_t]+b_f) \\
    & i_t = \sigma(W_i\cdot [h_{t-1}, x_t]+b_i) \\
    & \tilde{C_t} = \tanh(W_c\cdot[h_{t-1}, x_t]+b_i) \\
    & C_t = f_t * C_{t-1}+i_t * \tilde{C_t} \\
    & o_t = \sigma(W_o\cdot [h_{t-1}, x_t]+b_o) \\
    & h_t = o_t * \tanh(C_t) \\
  \end{split}
\end{equation}

Operations in LSTM include matrix multiplication, elementwise multiplication and activation computation. Quantization of those operators follow the equation in \ref{eq:quant_ops}. The value of bias involved in a linear transform is not quantized (see equation \ref{eq:quant_ops})
due to its negligible number of parameters compared to that of weight matrices

\begin{equation}
  \label{eq:quant_ops}
  \begin{split}
    & \v y = \v W\v x+\v b \rightarrow \v y = Q_n(\v W)Q_n(\v x)+\v b \\
    & \v y = \v x_1 *\v x_2 \rightarrow \v y = Q_n(\v x_1)*Q_n(\v x_2) \\
    & \v y = F(\v x) \rightarrow \v y = Q_n(F(\v x)) \\
    & F: \sigma, \tanh \\
  \end{split}
\end{equation}

We quantize every operation of LSTM in equation \ref{eq:quant_ops}, which means both input and parameters of any operator are quantized. Though quantization of input does not have any impact on the storage size of the model, computation can be accelerated because arithmetic with lower bit-width is faster. 
Lower bit-widths also mean we can squeeze more data into the same caches/registers which can reduce the frequency of accessing RAM, which usually consumes a lot of time and power.
In practice the overhead of value quantization (equation \ref{eq:quantization}) is negligible and we will detail on this later in section \ref{sec:qt_effect}.

\section{Experiments}
\label{sec:experiment}

\subsection{Experimental Setting}

\textbf{Data} The dataset we use is a subset from Audioset \cite{audioset}, which contains a large amount of 10-second audio clips. In particular, we select
dog sound, baby crying and gunshots as the target events. These three events included in Audioset amount to 13,460, 2,313 and 4,083
respectively, and we use all of them. 
In addition to these three events, we randomly selected 36,036 examples from all other audio clips in Audioset as negative samples. 
We randomly split the whole subset for training (70\%), validation (10\%) and test (20\%). Additional efforts has been made to 
ensure the distribution of events roughly the same across different sets.

\textbf{Implementation details} We compute log mel-filter bank energy (LFBE) features for each audio clip, as is common in previous works on AED \cite{many_cnns,cnn_augment,cakir2017}. It is calculated with window size 
of 25 ms and hop size of 10 ms at sampling rate of 16K. The number of mel coefficients is 64, which gives us log-mel spectrogram feature of size $998\times 64$ for each audio clip. 
Features are further normalized via global cepstral mean and variance normalization (CMVN).

We use DenseNet \cite{densenet} with 63 layers as the teacher model. The DenseNet we use contains 4 dense blocks with respectively 3, 6, 12 and 8 dense layers, where each layer is composed of batch normalization, ReLU, $1\times 1$ convolution, batch normalization, ReLU and $3\times 3$ convolution. We apply average pooling over all timesteps before feeding into a fully connected layer for classification. 
We experiment with different CNN architectures including ResNet \cite{resnet} and DenseNet with different number of layers (up to 121 layers). The DenseNet-63 gives us the best performance on validation set and we use it as the backbone teacher model.  

As mentioned in previous section, one-layer LSTM with 256 hidden units is used as the student model. Only the hidden state of last timestep is used for prediction.

In quantization experiments, we will experiment 8-bit and 4-bit quantization with training. We did not quantize the model with less than 4-bit, as it results in significant performance degradation. Both parameters and input are quantized to the target number of bit (8 or 4) except that the cell state ($C_t$ in equation \ref{eq:lstm}) will be quantized into 16 bits. 
Empirically we find quantizing $C_t$ to low-bits leads to divergence in training. This may be related to its potential unbounded value. Such behavior on quantizing recurrent neural networks is also reported in \cite{quant_rnn}.

For all experiments we use Adam optimizer with learning rate of 0.001 and batch size of 64. We tuned penalty on positive loss ($w_c$ in equation \ref{eq:sup_loss}) on validation set and found setting it to be the ratio between positive and negative examples of each class in the training set gives overall best results. This practice also prevents us from tuning $w_c$ for every class.

\textbf{Evaluation Metric} We evaluate the performance of models based on AUC (area under curve) on ROC curve (true positive rate vs. false positive rate) and EER (equal error rate) on DET curve (false negative rate vs. false positive rate). Higher AUC and lower EER indicate better performance. We report their values on individual events as well as average over all three events.

To evaluate the compression effect, we measure number of parameters (in Millions), size of parameters (in MB) and number of floating point operations (FLOPs). FLOPs measure the amount of computation while the first two metrics measure storage size of the model.

\subsection{Results}
\subsubsection{Effect of knowledge distillation}
\label{sec:kd_effect}
In this part, we will show: (1). LSTMs are more compact models but much less accurate compared to a deep CNN. (2). Accuacy of LSTM can be much improved through knowledge distillation from a deep CNN. Size of different models are listed in table \ref{tab:model_size}. As can be seen from table \ref{tab:model_size}, 1-layer LSTM is around $6$ times smaller compared to the DenseNet with 63 layers. However, Dense-63, which is more computational intensive, has achieved much higher accuracy than recurrent models (see table \ref{tab:kd}). The better performance is due to its depth as well as the convolutional architecture. By comparing LSTM of 1 and 3 layers, we find increasing depth of LSTM brings small gain but its performance still greatly falls behind CNN model. The low performance of LSTM is related to its recurrent architecture, where gradient vanishing will be severe when sequence is very long ($\approx$ 1000 frames). However, the gain of convolutional model can be distilled into LSTM despite their fundamental difference in architecture. Through knowledge distillation, error rates of one-layer LSTM are reduced consistently across all three events and on average EER is reduced by 26.7\% (relative).
The distilled 1-layer LSTM outperforms vanilla 3-layer LSTM, which shows the gain from knowledge distillation cannot be brought by simply increasing depth. In the meantime, the distilled LSTM is $3$ times smaller than its 3-layer counterpart which again shows knowledge distillation is an efficient way for model compression in AED.

\begin{table}[htp]
\centering
\small
\setlength{\tabcolsep}{3pt}
\caption{\label{tab:kd} AUC and EER of \textbf{full-precision} models (w and w/o knowledge distillation). For AUC, higher is better. For EER, lower is better}
\vspace{0.05in}
\begin{tabular}{c|c|cccc}\hline
 \multicolumn{2}{c|}{\begin{tabular}{c}Full-precision \\ models \end{tabular}} & Dense-63 & \begin{tabular}{@{}c@{}}LSTM \\ (3 layers) \end{tabular} & \begin{tabular}{@{}c@{}}LSTM \\ (1 layer) \end{tabular} & \begin{tabular}{@{}c@{}} LSTM+KD \\ (1 layer) \end{tabular} \\ \hline
\multirow{4}{*}{\begin{tabular}{c} AUC \\ (\%) \end{tabular}} & Dog & 95.68 & 92.51 & 91.35 & 93.65 \\
 & Baby & 97.80 & 93.66 & 92.09 & 95.81 \\
 & Gunshot & 97.93 & 94.90 & 92.86 & 96.78 \\ \cline{2-6}
 & \bf Avg & \textbf{97.14} & 93.87 & 92.10 & 95.41  \\ \hline
\multirow{4}{*}{\begin{tabular}{c}EER \\ (\%)\end{tabular}} & Dog & 11.11 & 15.19 & 16.60 & 14.07 \\
 & Baby & 6.56 & 13.58 & 15.58 & 10.21 \\
 & Gunshot & 6.41 & 11.79 & 13.07 & 9.19 \\ \cline{2-6}
 & \bf Avg & \textbf{8.03} & 13.52 & 15.08 & 11.16 \\ \hline
\end{tabular}
\end{table}

\begin{table}[htp]
\centering
\small
\setlength{\tabcolsep}{2pt}
\caption{\label{tab:model_size} Comparison of full-precision teacher and student models on number of parameters, parameter size and FLOPs}
\begin{tabular}{c|cccc}
 & Dense-63 & \begin{tabular}{c} LSTM \\ (3 layers) \end{tabular} & \begin{tabular}{c} LSTM \\ (1 layer) \end{tabular} & \begin{tabular}{c}LSTM+KD \\ (1 layer)\end{tabular} \\ \hline
 \# params (M) & 2.28 & 1.32 & \textbf{0.33} & \textbf{0.33} \\ \hline
 Param size(MB) & 8.70 & 5.27 & \textbf{1.26} & \textbf{1.26} \\ \hline
 FLOPs (G) & 1.38 & 1.31 & \textbf{0.32} & \textbf{0.32} \\ \hline
\end{tabular}
\end{table}

\subsubsection{Effect of quantized distillation}
\label{sec:qt_effect}
Quantizing the student LSTM can bring further compression effects.
We use the same 1-layer LSTM trained with knowledge distillation as in table \ref{tab:kd} except it is further quantized here.
AUC/EER of LSTM with different number of bits are shown in table \ref{tab:kd_qt_res}.
Oveall, quantization further reduces size of LSTM while maintaining its accuracy well.
LSTM can be compressed to $1/4$ and $1/8$ of full-precision size when it is quantized to 8 and 4 bits.
Relative EER increase under the two settings are respectively $4\%$ and $12\%$. As knowledge distillation improves model accuracy substantially, we find even a 4-bit quantized distilled model achieves much lower error rates compared to its full-precision counterpart without distillation (one-layer LSTM in table \ref{tab:kd}). Note result of 16-bit quantization is not shown in table \ref{tab:kd}. Under 16-bit setting, we find there is almost no loss of accuracy from quantization.

We do not measure FLOPs in table \ref{tab:kd_qt_res} because model parameters are quantized into integers and FLOPs no longer applies. However, we expect computation will be accelerated since we quantize every operation and lower bit-widths allows faster arithmetic computation. Detailed reason can be found in section \ref{sec:method}. Besides, the computation overhead brought by quantization is low. In our case, value quantization induces additional 0.0016G FLOPs, which is negligible compared to the 0.32G FLOPs of running full-precision LSTM. The exact speedup ratio by quantization depends implementation which are harware specific and will not be discussed here.

\begin{table}[htp]
\centering
\small
\caption{\label{tab:kd_qt_res} Performance of \textbf{quantized distilled} LSTM. For AUC, higher is better. For EER, lower is better}
\begin{tabular}{c|c|ccc}
  \multicolumn{2}{c|}{\begin{tabular}{c}LSTM+KD\\(1 layer)\end{tabular}} & \begin{tabular}{c}Full-\\precision\end{tabular} & 8-bit & 4-bit \\ \hline
\multirow{3}{*}{AUC (\%)} & Dog & 93.65 & 92.99 & 92.84 \\
 & Baby & 95.81 & 95.00 & 94.35 \\
 & Gunshot & 96.78 & 96.03 & 95.65 \\ \cline{2-5}
 & \bf Avg & \textbf{95.41} & 94.67 & 94.28 \\ \hline
\multirow{3}{*}{EER (\%)} & Dog & 14.07 & 14.61 & 14.72 \\
 & Baby & 10.21 & 10.19 & 12.56 \\
 & Gunshot & 9.19 & 10.23 & 10.31 \\ \cline{2-5}
& \bf Avg & \textbf{11.16} & 11.68 & 12.53 \\ \hline
 \multicolumn{2}{c|}{Param size(MB)} & 1.26 & 0.32 & \textbf{0.17} \\ \hline
\end{tabular}
\end{table}

\subsubsection{Further analysis on quantization}
\label{sec:analysis_qt}
In our quantization scheme the model is quantized both at training and test time. Quantization in training aims at learning proper range for parameters where the degradation from low-precision within that range is small. As both inputs and model parameters are quantized, the loss of computation precision is larger compared to the case where only parameters are quantized. Thus quantization in training is crucial to prevent degradation of performance. To further show the effectiveness of quantization in training, we compare it with post-mortem (PM) quantization where model is only quantized at test time. In addition we vary the number of hidden units of LSTM to see how performance gap changes with different model size. For simplicity we only show the average AUC/EER over three events but results are consistent acorss three events.

As can be seen from table \ref{tab:analysis}, quantization in training maintains accuracy better than PM in all scenarios. In particular, the performance gap is increasing as we lower bit-width.  
For instance of H=256, the relative degradation from full-precision model for ours and PM are $4.6\%$ and $5.7\%$ (in EER) in 8-bit setting. However, the degradation of PM increases drastically to $27.0\%$ in 4-bit setting while degradation of quantization in training is only $12.3\%$. Therefore, the lower bit-width model is, the more effective our approach becomes. 

In addition, we find that the gap between our quantized training scheme and PM grows as model scales down.
Similarly we consider relative degradation from full-precision model.
Both at 8-bit setting, difference of relative degradation between two quantization schemes increases from 1.1\% to 8.5\% in EER when number of hidden units is reduced from 256 to 64. Similar finding holds for 4-bit setting as well. Thus to an originally small model, quantization in training is important to prevent the accuracy drop when it is quantized. 

\begin{table}[htp]
\centering
\small
\setlength{\tabcolsep}{4pt}
\caption{\label{tab:analysis} Comparison between quantization in training and post-mortem for different model sizes. H: Number of hidden units of LSTM. Ours: quantization in training. PM: post-mortem, quantization at test-time only. For AUC, higher is better. For EER, lower is better. Note all models are trained with knowledge distillation}
\vspace{0.05in}
\begin{tabular}{c|ccccc}\hline
 (a). AUC(\%) & \begin{tabular}{@{}c@{}}Full\\ precision\end{tabular} & \begin{tabular}{@{}c@{}}Ours \\ 8-bit\end{tabular} & \begin{tabular}{@{}c@{}}PM \\ 8-bit\end{tabular} & \begin{tabular}{@{}c@{}}Ours \\ 4-bit\end{tabular} & \begin{tabular}{@{}c@{}}PM \\ 4-bit\end{tabular} \\ \hline
            H=256 & 95.41 & 94.67 & 94.55 & 94.28 & 92.64 \\ \hline
            H=128 & 94.92 & 94.61 & 94.47 & 93.98 & 91.40 \\ \hline
            H=64 & 94.30 & 94.04 & 93.41 & 93.10 & 89.52 \\ \hline\hline
            (b). EER(\%)  & \begin{tabular}{@{}c@{}}Full\\ precision\end{tabular} & \begin{tabular}{@{}c@{}}Ours \\ 8-bit\end{tabular} & \begin{tabular}{@{}c@{}}PM \\ 8-bit\end{tabular} & \begin{tabular}{@{}c@{}}Ours \\ 4-bit\end{tabular} & \begin{tabular}{@{}c@{}}PM \\ 4-bit\end{tabular} \\ \hline
                      H=256 & 11.16 & 11.68 & 11.80 & 12.53 & 14.17 \\ \hline
                      H=128 & 11.28 & 11.53 & 12.08 & 12.63 & 16.57 \\ \hline
                      H=64 & 12.05 & 12.24 & 13.27 & 12.97 & 17.95 \\ \hline
\end{tabular}
\end{table}

\section{Conclusion}
We study the model compression problem in the context of acoustic event detection. Our compression scheme jointly applies knowledge distillation and quantization to the target model. Experimental results show that the performance of shallow LSTM model can be greatly improved via knowledge distillation without increase of size. The distilled model can be further compressed to an order of magnitude smaller by quantization in training while the accuracy is well preserved. Analysis on multiple student models further verify the robustness of our proposed method, as well as its advantage over naive quantization scheme.

\bibliographystyle{IEEEtran}

\bibliography{refs}

\end{document}